\documentstyle[epsfig,aps]{revtex}
\def\bge{\begin{equation}}
\def\ene{\end{equation}}
\def\bg{\begin{eqnarray}}
\def\en{\end{eqnarray}}

\def\ubar{{\bar{u}}}
\def\dbar{{\bar{d}}}
\def\sbar{{\bar{s}}}

\def\vr{\vec{r}}

\def\e{\epsilon}
\def\pomega{{\omega_B}}
\begin{document}
\def\bra#1{{\langle #1{\left| \right.}}}
\def\ket#1{{{\left.\right|} #1\rangle}}
\def\bfgreek#1{ \mbox{\boldmath$#1$}}
\title{Recent results from QMC relevant to TJNAF} 

\author{A.W. Thomas, D.H. Lu, K. Tsushima and A.G. Williams}
\address{Department of Physics and Mathematical Physics and \break
        Special Research Centre for the Subatomic Structure of Matter,\break
        University of Adelaide, Australia 5005}
\author{K. Saito}
\address{Physics Division, Tohoku College of Pharmacy,\break
Sendai 981-8558, Japan}
\maketitle

\vspace{-6.9cm}
\begin{flushright}
{\footnotesize Invited talk at the TJNAF Users Workshop} \\
{\footnotesize Newport News, June 15-18, 1998} \\
{\footnotesize ADP-98-36/T309 \hspace{2cm}}
\end{flushright}
\vspace{5.9cm}
\begin{abstract}
We review recent calculations using the quark meson coupling model which
should be of particular interest at Jefferson Lab. In particular, we
discuss the change in the proton electric and magnetic form factors when
it is bound in a specific shell model orbit. Modern quasi-elastic
electron scattering experiments should be able to detect effects of the
size predicted. We also examine the mean field potential felt by an
$\omega$ in a finite nucleus, concluding that the $\omega$ should be
bound by between 50 and 100 MeV.
\end{abstract}

\section{Introduction} 

Whether or not quark degrees of freedom play a significant role 
in nuclear theory is one of the most fundamental
questions in strong interaction  physics.
Tremendous efforts have been devoted to the study of medium modifications 
of hadron properties\cite{matter95}.
The idea that nucleons might undergo considerable change of their internal 
structure in a baryon-rich environment is supported by our understanding
of asymptotic freedom in QCD. In particular, at sufficiently high
density one expects a transition to a new phase of matter where quarks
and gluons are deconfined and it seems unlikely that there would be no
indication of this below the critical density. There has also been quite
a bit of encouragement from 
experiment, including the discovery of 
the variation of nucleon structure 
functions in lepton deep-inelastic scattering off nuclei 
(the nuclear EMC effect)\cite{EMC}, 
the quenching of the axial vector coupling constant, $g_A$, 
in nuclear $\beta$-decay\cite{gA}, and the missing strength 
of the response functions in nuclear quasielastic  electron 
scattering\cite{quasi}. 

There have been several effective Lagrangian approaches in the literature
dealing with modifications of the nucleon size and electromagnetic
properties in medium\cite{Meissner89,medium}.
All of these investigations have found that 
nucleon electromagnetic form factors
are suppressed and the rms radii of the proton somewhat increased in
bulk nuclear matter.
In Ref.\cite{medium}, we examined medium modifications of nucleon
electromagnetic properties in nuclear matter, using the quark-meson
coupling model (QMC)\cite{Guichon,finite}.
The self-consistent change in the internal structure of a bound nucleon
was found to be
consistent with the constraints from $y$-scaling data\cite{sick}
and the Coulomb sum rule\cite{coulomb}.
Here we present a preliminary report on our investigations of the 
electromagnetic form factors for a nucleon bound in specific, shell
model orbitals of realistic, finite nuclei.
This is of direct relevance to quasielastic scattering measurements
underway at TJNAF\cite{TJNAF}.

One of the most obvious changes in a particle's in-medium properties is
its effective mass and 
the medium modification of the light vector ($\rho$, $\omega$ and
$\phi$) meson masses has been investigated extensively by
many authors~\cite{maru}--\cite{saitomega}.
It has been suggested that dilepton production in the nuclear
medium formed in relativistic heavy ion collisions, can provide a unique
tool to measure such modifications as meson mass shifts.
For example, the experimental data 
obtained at the CERN/SPS by the CERES~\cite{ceres} and HELIOS~\cite{hel}
collaborations has been interpreted as evidence for a downward shift of
the $\rho$ meson mass in dense nuclear matter~\cite{li}.
To draw a more definite conclusion,
measurements of the dilepton spectrum from vector
mesons produced in nuclei are planned at TJNAF~\cite{jlab} and
GSI~\cite{gsi}.

Recently, a new method to study meson mass shifts in
nuclei was proposed by Hayano {\it et al.}~\cite{hayano}.
Their suggestion is to use the (d, $^3$He) reaction to produce
$\eta$ and $\omega$ mesons with nearly zero recoil momentum.
If the meson feels a large enough, attractive (Lorentz scalar) force inside
a nucleus it will form
a meson-nucleus bound state -- of course, there is a cancellation of the
mean field vector potential for a quark-anti-quark pair. 
Hayano {\it et al.}~\cite{hayano2} estimated the binding energies
for various $\eta$-mesic nuclei. They also
calculated some quantities for the $\omega$ meson case. 
Their $\eta$-nucleus optical potential was calculated to first-order
in density, using the $\eta$-nucleon scattering length as input.
We recently investigated this problem \cite{qmcom}, using QMC 
to study whether it is possible to form
$\eta$- and/or $\omega$-nucleus bound states
in $^{16}$O, $^{40}$Ca, 
$^{90}$Zr and $^{208}$Pb, as well as
$^{6}$He, $^{11}$B and $^{26}$Mg. The latter three nuclei
correspond to the proposed experiment at GSI -- i.e., 
the reactions, $^7$Li\,(d,$^3$He)\,$^6_{\eta/\omega}$He,
$^{12}$C\,(d,$^3$He)\,$^{11}_{\eta/\omega}$B and
$^{27}$Al\,(d,$^3$He)\,$^{26}_{\eta/\omega}$Mg.
Here we shall briefly report our results for the $\omega$ case, which is
again of direct interest at Jefferson Lab.

\section{The Quark Meson Coupling Model}

The details of the derivation of the  Quark Meson Coupling 
model (QMC) for finite nuclei may be found
in Ref.~\cite{finite}. Here we briefly summarize the essential points.
At least as far as the single particle energies are concerned,
the QMC model for spherical finite nuclei, in mean-field approximation, 
can be summarized in an effective Lagrangian density\cite{finite}
\begin{eqnarray}
{\cal L}_{QMC}&=& \overline{\psi}({\vec r}) [i \gamma \cdot \partial 
- m_N + g_\sigma(\sigma({\vec r})) \sigma({\vec r})  
- g_\omega \omega({\vec r}) \gamma_0 \nonumber \\
&-& g_\rho \frac{\tau^N_3}{2} b({\vec r}) \gamma_0 
- \frac{e}{2} (1+\tau^N_3) A({\vec r}) \gamma_0 ] \psi({\vec r}) \nonumber \\
&-& \frac{1}{2}[ (\nabla \sigma({\vec r}))^2 + 
m_{\sigma}^2 \sigma({\vec r})^2 ] 
+ \frac{1}{2}[ (\nabla \omega({\vec r}))^2 + m_{\omega}^2 
\omega({\vec r})^2 ] \nonumber \\
&+& \frac{1}{2}[ (\nabla b({\vec r}))^2 + m_{\rho}^2 b({\vec r})^2 ] 
+ \frac{1}{2} (\nabla A({\vec r}))^2 , 
\label{qmclag}
\end{eqnarray}
where
$\psi({\vec r})$,$\sigma({\vec r})$, $\omega({\vec r})$, $b({\vec r})$, 
and $A({\vec r})$ are the nucleon, $\sigma$, $\omega$, $\rho$,
and Coulomb fields, respectively.
Note that only the time components of the $\omega$ and 
neutral $\rho$ fields are kept in the mean field approximation.
These five fields now depend on position $\vec{r}$, 
relative to the center of the nucleus.
The spatial distributions are determined by solving the equations of motion
self-consistently.

At the level of the effective Lagrangian density, the 
key difference between QMC  
and QHD\cite{QHD} lies only in the $\sigma NN$ coupling constant,
which depends on the scalar field strength in QMC 
while it remains  constant in QHD.  
The coupling constants $g_\sigma(0)$, $g_\omega$ and  $g_{\rho}$  
are fixed to reproduce the saturation properties and the bulk symmetry energy
of nuclear matter.
The only free parameter, $m_\sigma$, which controls the range of the 
attractive 
interaction, and therefore affects the nuclear surface slope and its thickness,
is fixed by fitting the experimental charge rms radius of $^{40}Ca$, while
keeping the ratio $g_\sigma/m_\sigma$ intact, as constrained 
by the nuclear matter properties. We note that one of the major
successes of the model in nuclear matter is that it always produces a
value for the nuclear incompressibility that is in reasonable agreement
with experiment.

\section{Form Factor Modifications in Medium}

Of course, the main difference between QMC and conventional treatments
of nuclear structure is that in QMC the internal structure of the
``nucleon'' self-consistently adjusts to the local medium in which it
sits.
The quark wave function, as well as the nucleon wave function
(both are Dirac spinors), are determined 
once a solution to equations of motion are found.
The electromagnetic form factors for a proton bound in a specific
orbit $\alpha$, 
in  local density approximation,
are  simply given by
\begin{equation}
G_{E,M}^\alpha(Q^2)= \int G_{E,M} (Q^2,\rho_B(\vec{r})) 
\rho_{p\alpha}(\vec{r}) \,d\vec{r},
\end{equation}
where $G_{E,M}(Q^2,\rho_B(\vec{r}))$
is  the density-dependent form factor of a ``proton'' immersed in 
nuclear matter with a local baryon density, $\rho_B(\vec{r})$.
Using the calculated nucleon shell model wave functions, 
the local  baryon density and the local proton density in the specified
 orbit $\alpha$, are easily evaluated. 

The notable medium modifications of the quark wavefunction inside the bound 
``nucleon'' in QMC include a reduction of its frequency and an enhancement 
of the lower component of the Dirac spinor.
As in earlier work, the corrections arising from recoil and center of mass 
motion for the bag are made  using the Peierls-Thouless
projection method, combined with  Lorentz contraction of the internal
quark wave function, and   
the perturbative pion cloud is added afterwards\cite{ltw}.
Additional off-shell form factors and possible 
meson exchange currents have been omitted 
in the present, exploratory investigation.  
The resulting nucleon electromagnetic form factors agree with  experiments
quite well in free space\cite{ltw}, at least for momentum transfers less
than 1 $\mbox{GeV}^2$. (The region of validity is mainly constrained by 
limitations of the bag model.)  
In order to reduce the theoretical uncertainties at higher momentum
transfer, which is also of experimental interest, we prefer to show the ratios
of the form factors with respect to corresponding free space values.

\begin{figure}
\centering{\
\epsfig{file=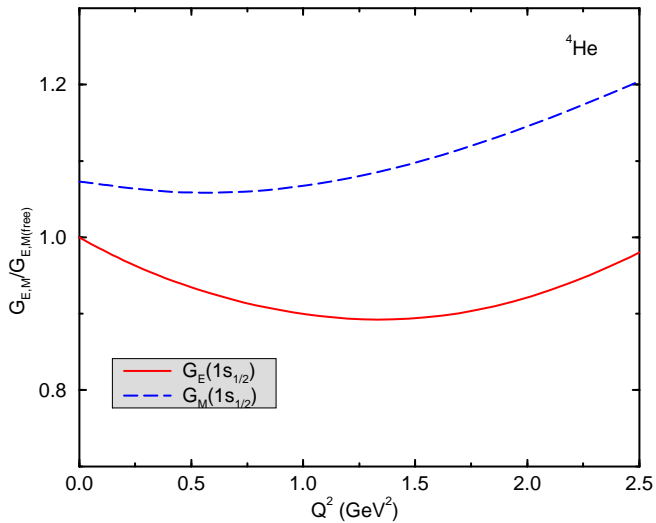,width=10cm}
\caption{Ratio of in-medium to free space electric and
magnetic form factors of the proton in $^4He$.
(The free bag radius was taken to be $R_0=0.8$ fm.)}
\label{he4.ps}}
\end{figure}
\begin{figure}
\centering{\
\epsfig{file=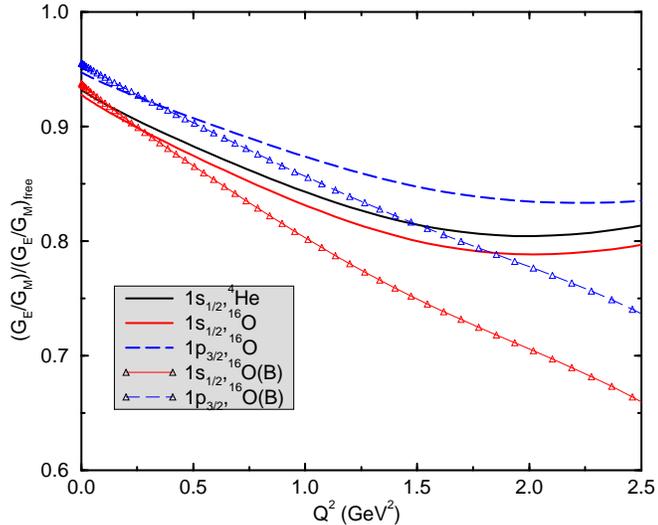,width=10cm}
\caption{Ratio of electric and magnetic form factors in-medium divided
by free space ratio.
As in previous figure, curves with triangle symbols represent
corresponding
values calculated in a variant of QMC with a 10\% reduction of $B$.}
\label{2ratio.ps}}
\end{figure}

Fig.~\ref{he4.ps} shows the ratio of the electric and magnetic form factors  
with respect to the free space values
for $^4He$ (only one state, $1s_{1/2}$).
As expected, both the electric and magnetic rms  radii become slightly larger
and the magnetic moment of the proton increases by about 7\%.
Although we cannot show the results for lack of space, we have made
similar calculations for $^{16}O$, where the 
momentum dependence of the form factors
for the $1s$-orbit nucleon is  more supressed than that for the
$1p$-states. This is because the inner orbit 
experiences a larger average baryon density.
The magnetic moment of the nucleon in the $1s$-orbit 
is similar to that in $^4He$, 
but it is reduced  by 2--3\% in the  $1p$-orbits.
The difference between two $1p$-orbits is rather small.

{}From the experimental point of view, we note that the ratio 
$G_E/G_M$ can be derived directly from the ratio of transverse 
to longitudinal polarization of the outgoing proton,
with minimal systematic errors. 
We find that $G_E/G_M$ (for a proton in $^4He$)  runs roughly 
from 0.41 at $Q^2 = 0 $ to 0.28 at $Q^2 = 1 \mbox{ GeV}^2$.
The  ratio of  $G_E/G_M$ with respect to the corresponding free
space ratio is presented in Fig.~\ref{2ratio.ps}.
The result for the $1s$-orbit  in $^{16}O$ is close to that in  $^4He$ 
and   2\% lower than
that for the $p$-orbits in $^{16}O$.
In order to interpolate smoothly between the confined and deconfined
phases as the baryon density increases, it has been suggested that the
bag constant might decrease with increasing density. As we see in 
Fig.~\ref{2ratio.ps} the 
effect of a possible reduction in $B$ has a significant effect
on this ratio of ratios, especially for larger $Q^2$.

{}For completeness, we have also calculated the orbital electric and magnetic 
form factors for heavy nuclei such as $^{40}Ca$ and $^{208}Pb$.
Because of the larger central baryon density of heavy nuclei, 
the proton electric and magnetic form factors 
in the inner orbits ($1s_{1/2}$, $1p_{3/2}$ and $1p_{1/2}$ orbits) 
suffer much stronger medium modifications
than those in light nuclei.
That is to say, the $Q^2$ dependence is further suppressed, while
the magnetic moments appear to be larger.
Surprisingly, the nucleons in peripheral orbits ($1d_{5/2}$, $2s_{1/2}$, 
and $1d_{3/2}$ for $^{40}Ca$
and $2d_{3/2}$, $1h_{11/2}$, and $3s_{1/2}$ for $^{208}Pb$) 
still endure significant medium effects ($1d_{3/2}$ overlaps with $1d_{5/2}$  
and $3s_{1/2}$ overlaps with $2d_{3/2}$), and
are comparable to those in $^4He$.

\section{Omega-mesic nuclei}

As we explained in the introduction there is great interest in
the medium modification of the light vector ($\rho$, $\omega$ and
$\phi$) meson masses.
In earlier work with the QMC model we have investigated changes of 
meson properties in the nuclear medium -- notably, the effective mass of the 
$\rho$ meson formed in light nuclei 
and properties of the kaons 
in nuclear matter~\cite{qmck}.
However, we have only recently tackled the question of whether   
meson-nucleus bound states are possible in the model \cite{qmcom}.  

At position $\vr$ in a nucleus (the coordinate origin is taken  
as the center of the nucleus), 
the Dirac equations for the quarks and antiquarks 
in the $\omega$ meson bag are given 
by~\cite{qmck}:
\bge
\left[ i \gamma \cdot \partial - (m_q - V_\sigma(\vr)) \mp \gamma^0
\left( V_\omega(\vr) + \frac{1}{2} V_\rho(\vr) \right) \right]
\left(\begin{array}{c} \psi_u(\vr)\\ \psi_\ubar(\vr)\\ \end{array}\right) 
 = 0,
\label{diracu}
\ene
\vspace{-0.5cm}
\bge
\left[ i \gamma \cdot \partial - (m_q - V_\sigma(\vr)) \mp \gamma^0
\left( V_\omega(\vr) - \frac{1}{2} V_\rho(\vr) \right) \right]
\left(\begin{array}{c} \psi_d(\vr)\\ \psi_\dbar(\vr)\\ \end{array} \right) 
 = 0,
\label{diracd}
\ene
\vspace{-0.5cm}
\bge
\left[ i \gamma \cdot \partial - m_s \right] 
\psi_s (\vr)\,\, ({\rm or}\,\, \psi_\sbar(\vr)) = 0,
\label{diracs}
\ene
where $V_\sigma(\vr) = g^q_\sigma \sigma(\vr), 
V_\omega(\vr) = g^q_\omega \omega(\vr)$ and
$V_\rho(\vr) = g^q_\rho b(\vr)$ are 
the mean-field potentials at the position $\vr$, and  are calculated 
self-consistently, as we explained earlier.
Hereafter we use the notation, $\pomega$, to specify the physical,  
bound $\omega$ meson, in order to avoid confusion
with the isoscalar-vector $\omega$ field appearing in QMC (or QHD).
The bag radius in medium, $R^*$, is determined self-consistently
through the stability condition for the (in-medium) mass of the meson 
against the variation of the bag radius.

The eigenenergies in units of $1/R^*$, $\epsilon_f\;
(f=u,\ubar,d,\dbar)$, are given by
\begin{eqnarray}
\left( \begin{array}{c} \e_{u}(\vr) \\ \e_{\ubar}(\vr) \end{array} \right)
&=& \Omega_q^*(\vr) \pm R^* \left(
V_\omega(\vr) + \frac{1}{2} V_\rho(\vr) \right), \\
\left( \begin{array}{c} \e_{d}(\vr) \\ \e_{\dbar}(\vr) \end{array} \right)
&=& \Omega_q^*(\vr) \pm R^* \left(
V_\omega(\vr) - \frac{1}{2} V_\rho(\vr) \right),
\label{quarkenergy}
\en
where $\Omega_q^*(\vr) = \sqrt{x_q^2 + (R^* m_q^*)^2}$, with
$m_q^* = m_q - g^q_\sigma \sigma(\vr)\, (q=u,\ubar,d,\dbar)$.
The bag eigenfrequencies, $x_q$, are 
determined by the usual, linear boundary condition~\cite{qmck}. 

The physical $\pomega$ meson is a 
superposition of the octet and singlet states with the mixing angle, 
$\theta_V = 39^\circ$, as estimated by Particle Data Group.
We assume that the value of the mixing angle does not change in 
medium, although this is possible and merits further investigation.
We self-consistently  
calculate the effective mass, $m^*_\pomega(\vr)$  
at the position $\vr$ in the nucleus. 
Because the vector potentials for 
the same flavor of quark and antiquark cancel each other,  
the potentials for the $\pomega$ meson is given by
$m^*_\pomega(r) - m_\pomega$, 
which depends only on the distance from the center of the 
nucleus, $r = |\vr|$.

The depth of the potential felt by the $\omega$ meson is  
typically more than 100 MeV.  
Because the typical momentum of the bound $\omega$ is low, it should be 
a very good approximation to neglect the possible energy difference 
between the longitudinal and transverse components of the 
$\omega$~\cite{saitomega}.
Imposing the Lorentz condition, $\partial_\mu \phi^\mu = 0$,  
solving the Proca equation becomes equivalent to solving the Klein-Gordon
equation. Thus, to obtain the meson nucleus binding energies, 
we may solve the Klein-Gordon equation. 

An additional complication, which has so far been ignored, is the 
meson absorption in the nucleus. This requires 
an imaginary part for the potential to describe the effect. 
At the moment, we have not been able to calculate the in-medium widths of 
the mesons, or the imaginary part of the potential in medium, 
self-consistently within the model. 
In order to make a more realistic estimate for the meson-nucleus bound states,
we therefore include the width of the $\pomega$ meson in the nucleus 
phenomenologically: 
\bg
\tilde{m}^*_\pomega(r) &=&
m^*_\pomega(r) - \frac{i}{2} 
\left[ (m_\pomega - m^*_\pomega(r)) 
\gamma_\pomega + \Gamma_\pomega \right],
\label{imaginary}\\
&\equiv& m^*_\pomega(r) - \frac{i}{2} \Gamma^*_\pomega (r),
\label{width}
\en
where, $m_\pomega$ and $\Gamma_\pomega$ is the corresponding mass and width 
in free space.
In Eq.~(\ref{imaginary}) $\gamma_\pomega$ is treated as a phenomenological 
parameter to describe the in-medium meson width,
$\Gamma^*_\pomega(r) \equiv
(m_\pomega - m^*_\pomega(r)) \gamma_\pomega + \Gamma_\pomega$.
According to the estimates in Refs.~\cite{fri,hayano}, 
the width of the $\pomega$ meson at normal nuclear matter density 
is $\Gamma^*_\pomega \sim 30 - 40$ MeV.
Thus, we calculate the single-particle energies using the values 
for the parameters appearing in Eq.~(\ref{imaginary}), 
$\gamma_\pomega = 0, 0.2, 0.4$, which cover the estimated ranges. 
Thus we actually solve the following Klein-Gordon equation:
\bge
\left[ 
\nabla^2 + E^2_\pomega - \tilde{m}^{*2}_\pomega(r) 
\right]\, \phi_\pomega(\vr) = 0. 
\label{kgequation2}
\ene

Equation~(\ref{kgequation2}) has been solved in momentum 
space, using the method developed in Ref.~\cite{landau}.
To confirm the calculated results, we also calculated the 
single-particle energies by solving 
the Schr\"{o}dinger equation.
Calculated single-particle energies for the $\pomega$ 
meson, obtained by solving the Klein-Gordon equation  
are listed in Table~\ref{omegaenergy}. 
We should mention that the advantage of solving the Klein-Gordon 
equation in momentum space is that it can handle quadratic terms
arising in the potentials without any trouble.

\begin{table}
\begin{center}
\caption{ 
Calculated $\omega$ meson single-particle energies, 
$E = Re(E_\omega - m_\omega)$, 
and full widths, $\Gamma$, (both in MeV) in various nuclei, where 
the complex eigenenergies are, $E_\omega = E + m_\omega - i \Gamma/2$. 
See Eq.~(\protect\ref{imaginary}) for 
the definition of $\gamma_\omega$. In the light of 
the estimates of $\Gamma$ in 
Refs.~\protect\cite{fri}, the results with $\gamma_\omega = 0.2$ are
expected to correspond best with experiment.
}
\vspace{0.25cm}
\label{omegaenergy}
\begin{tabular}[t]{lc|cc|cc|cc}
& &$\gamma_\omega$=0 & &$\gamma_\omega$=0.2& &$\gamma_\omega$=0.4& \\
\hline \hline
& &$E$ &$\Gamma$ &$E$ &$\Gamma$ &$E$ &$\Gamma$ \\
\hline
$^{16}_\omega$O &1s &-93.5&8.14 &-93.4&30.6 &-93.4&53.1 \\
                &1p &-64.8&7.94 &-64.7&27.8 &-64.6&47.7 \\
\hline
$^{40}_\omega$Ca &1s &-111&8.22  &-111&33.1  &-111&58.1 \\
                 &1p &-90.8&8.07 &-90.8&31.0 &-90.7&54.0 \\
                 &2s &-65.6&7.86 &-65.5&28.9 &-65.4&49.9 \\
\hline
%
$^{90}_\omega$Zr &1s &-117&8.30  &-117&33.4  &-117&58.6 \\ 
                 &1p &-105&8.19  &-105&32.3  &-105&56.5 \\  
                 &2s &-86.4&8.03 &-86.4&30.7 &-86.4&53.4 \\   
\hline
$^{208}_\omega$Pb &1s &-118&8.35 &-118&33.1 &-118&57.8 \\ 
                  &1p &-111&8.28 &-111&32.5 &-111&56.8 \\ 
                  &2s &-100&8.17 &-100&31.7 &-100&55.3 \\ 
\hline \hline
$^{6}_\omega$He &1s &-55.7&8.05 &-55.6&24.7 &-55.4&41.3 \\
\hline
$^{11}_\omega$B &1s &-80.8&8.10 &-80.8&28.8 &-80.6&49.5 \\
\hline
$^{26}_\omega$Mg &1s &-99.7&8.21 &-99.7&31.1 &-99.7&54.0 \\
                 &1p &-78.5&8.02 &-78.5&29.4 &-78.4&50.8 \\
                 &2s &-42.9&7.87 &-42.8&24.8 &-42.5&41.9 \\
%
%
\end{tabular}
\end{center}
\end{table}
%
%
%

Our results strongly support the suggestion of Hayano {\it et al.}
that one should find  
bound $\omega$-nuclear states \cite{hayano,hayano2}. For 
a large atomic number nucleus and a relatively 
wide range of the in-medium meson widths, it seems inevitable that one  
should find such $\omega$-nucleus bound states, bound by 50-100 MeV. 
{}For a more consistent treatment, 
we would like to calculate the in-medium meson width  
within the QMC model self-consistently.
We would also need to take into account the $\sigma$-$\omega$ 
mixing effect which is very interesting, and especially important 
at higher densities~\cite{saitomega}. 

\section{Conclusion}

In summary, we have calculated the electric and magnetic form factors
for the proton bound in  specific orbits for several
closed shell finite nuclei.
Generally the electromagnetic rms radii and the magnetic moments of the
bound proton are increased by the medium modifications.
The form factors corresponding to different orbits appear to behave
quite differently.
While the difference between the nucleon form factors for orbits split
by the spin-orbit force is very small, the difference between inner and
peripheral orbits is considerable.
In view of current experimental developments, including the ability to
precisely measure
elastic and quasielastic electron-nucleus scattering polarization
observables, it should be possible to detect differences between the
form factors in different shell model orbits.
The current and future experiments at
TJNAF and Mainz therefore promise to provide vital information
with which to guide and constrain dynamic
microscopic models for finite  nuclei, and perhaps unambiguously
isolate
a signature for the role of quarks.

We have also calculated the single-particle energies for  
$\eta$- and $\omega$-mesic  nuclei using QMC.
Here we reported only the results for the $\omega$.
The potentials for the mesons in the nucleus were calculated 
self-consistently, in local density approximation, by embedding the 
MIT bag model $\omega$ meson in the nucleus described 
by solving mean-field equations of motion. 
Although the specific form for the width of the in-medium  meson
could not be calculated in this model, 
our results suggest that one should find  
deeply bound $\omega$-nucleus states for a relatively wide range 
of the in-medium meson widths.
In the near future, we plan to calculate the 
in-medium $\omega$ width self-consistently in the QMC model.
 
%
%
\vspace{0.5cm}

\noindent{\bf Acknowledgment}\\
This work was supported by the Australian Research Council.
%

%

\end{document}